\documentclass[twocolumn]{aastex63}
\usepackage{bm}

\shorttitle{RCR Implementation}
\shortauthors{Konz et al.}
\graphicspath{{./}{figures/}}

\begin{document}

\title{Robust Chauvenet Rejection: Powerful, but Easy to Use Outlier Detection for Heavily Contaminated Data Sets}

\author[0000-0003-0230-1598]{Nicholas C. Konz}
\affiliation{Department of Physics and Astronomy\\
University of North Carolina at Chapel Hill \\
Chapel Hill, NC 27599, USA}
\affiliation{Department of Electrical and Computer Engineering\\
Duke University \\
Durham, NC 27708, USA}

\author[0000-0002-5060-3673]{Daniel E. Reichart}
\affiliation{Department of Physics and Astronomy\\
University of North Carolina at Chapel Hill \\
Chapel Hill, NC 27599, USA}

\begin{abstract}
In \citet{maples2018robust} we introduced Robust Chauvenet Outlier Rejection, or RCR, a novel outlier rejection technique that evolves Chauvenet's Criterion by sequentially applying different measures of central tendency and empirically determining the rejective sigma value. RCR is especially powerful for cleaning heavily-contaminated samples, and unlike other methods such as sigma clipping, it manages to be both accurate and precise when characterizing the underlying uncontaminated distributions of data sets, by using decreasingly robust but increasingly precise statistics in sequence. For this work, we present RCR from a software standpoint, newly implemented as a Python package while maintaining the speed of the C++ original. RCR has been well-tested, calibrated and simulated, and it can be used for both one-dimensional outlier rejection and $n$-dimensional model-fitting, with or without weighted data. RCR is free to use for academic and non-commercial purposes, and the code, documentation and accompanying web calculator can be found and easily used online at \url{https://github.com/nickk124/RCR}.











\end{abstract}



\section{Introduction} \label{sec:intro}
Data is everywhere, and nearly as ubiquitous as the data itself is the presence of outliers: measurements that are not part of the distribution that we wish to characterize. If we desire to (1) use these measurements to estimate a single value or (2) fit some model to the data, outliers can easily result in erroneous predictions.

Consider a sample of data points, possibly weighted, that contains both outlying and non-outlying measurements. The non-outliers are drawn from a given statistical distribution given some physical process, and the outliers are sample contaminants drawn from another distribution (often unknown), due to either a different physical process or non-statistical measurement errors.

As we described in \citet{maples2018robust}, hereafter RCR1, there are a host of methods for detecting and rejecting outliers, the most ubiquitous being sigma clipping, i.e., rejecting any measurement outside of some number of standard deviations away from the mean. Sigma clipping, however, has a number of downsides (see RCR1 \S 1), namely that measurements from the non-outlier distribution may be inadvertently rejected when using the technique, or conversely that not enough of the outliers are rejected (as how can we properly determine the number of sigmas to reject past?). One of the most straightforward improvements over sigma clipping is Chauvenet's criterion \citep{chavuenet1863manual}, which simply states that a measurement is deemed outlying if
\begin{equation} \label{eq:chauvenet}
    NP(\mathrm{>}|z|)<0.5,
\end{equation}
where $N$ is the number of measurements of the sample, and $P(\mathrm{>}|z|)$ is the cumulative probability of the measurement being more than $z$ standard deviations from the mean, assuming a Gaussian/normal distribution. 

Despite Chauvenet rejection (CR) being superior to sigma clipping, it still suffers from the fact that its defining parameters, the mean and the standard deviation, are not very \textit{robust} quantities, meaning that both can be easily contaminated by the very outliers that need to be rejected. As such, the first improvement that RCR makes over traditional CR is to utilize more robust versions of these statistics, shown in \S\ref{sec:robust}. RCR works by applying Chauvenet's criterion iteratively, rejecting some number of measurements in bulk upon each cycle\footnote{Rejecting in bulk can be almost as effective as rejecting one at a time, while being significantly more computationally efficient (see \S\ref{sec:bulk}).}. After each iteration, we lower $N$ by the number of rejected points, and re-estimate the mean and standard deviation (or their corresponding more-robust versions) in order to evaluate Equation \ref{eq:chauvenet}. We assume that the uncontaminated measurements are approximately Gaussian-distributed, but we do not necessarily make any assumptions about the distribution of the contaminants.

In \S\ref{sec:functional} we show how RCR can be applied to model-fitting, by generalizing the aforementioned single-dimensional rejection methods into $n$-dimensional counterparts. Next, in \S\ref{sec:results} we give a brief overview of a few applications of RCR, from the more ``textbook'' cases of one-dimensional (\S\ref{sec:results_value}) and model-fitting (\S\ref{sec:results_functional}) outlier rejection, to the more complex examples of aperture photometry (\S\ref{sec:results_aperture}) and a radio telescope observation mapping pipeline (\S\ref{sec:results_radio}).

In \S \ref{sec:discussion} we discuss various points to keep in mind when using RCR, as well as situations where RCR could fail, and we conclude with \S\ref{sec:conclusion}. Finally, in the Appendix we describe (1) how to install RCR and contribute to its future development, (2) the online documentation for the RCR package and (3) a web-based calculator through which RCR can be quickly and easily used.

\section{Methods} \label{sec:methods}

\subsection{Robust Techniques} \label{sec:robust} 
The three most commonly-used measures of central tendency, the mean, median and mode, are increasingly robust (i.e., sensitive to outliers), yet decreasingly precise (RCR1). We compute the mean and median in the canonical ways, while for the mode, we use the iterative half-sample method of \citet{bickel2005fast}, which is described on pg.\ 2 of RCR1. If the data is weighted, these statistics can be generalized using the prescription given in \S6 of RCR1. 

There are a variety of ways to estimate sample deviation, each of which can be paired with a metric of central tendency. The most used of these is the standard deviation, which can naturally be paired with the mean. In RCR1, we developed a more robust measure of sample deviation that in turn can be paired with the median and the mode, that we label as ``68.3-percentile deviation'', or $\sigma_{68.3}$. In RCR1 we defined $\sigma_{68.3}$ in three increasingly robust ways; in practice, however, we only use the first and third of these in RCR\footnote{This is true within the ``main'' loop of RCR, shown in Fig.\ \ref{fig:flow_techniques}. The second technique is used, however, for bulk pre-rejection (\S\ref{sec:bulk}, Fig.\ \ref{fig:flow_withbulk}); see Fig.\ 30 of RCR1.} (see Fig.\ \ref{fig:flow_techniques}), as we have found the techniques that utilize the second to be inferior. We also note that the following description of these techniques is for data that is unweighted/equally-weighted; refer to \S6 of RCR1 for the general case of weighted data.

The first method that we use to define the 68.3-percentile deviation is to sort the absolute values of the deviations of each measurement from the measure of central tendency in question $\mu$ (either the median or the mode), and then take the 68.3-percentile value from this sorted distribution. This is analogous to the median absolute deviation, or MAD, except that the MAD takes the 50-percentile value.

The third method used to define $\sigma_{68.3}$ is a bit more involved, and is performed as follows. Consider $N$ normally-distributed, equally-weighted measurements (we generalize this for other distributions as described in \S\ref{sec:nongauss}) that are sorted by the absolute value of their deviations from $\mu$. The percentile of the $i$th element is given by
\begin{equation} \label{68.3_equation1}
\frac{i-1+\Delta i}{N} = P\left(\rm{<}\left|\frac{\delta_i}{\sigma_{68.3}}\right|\right),
\end{equation}
where $P(\rm{<}|\delta_i/\sigma_{68.3}|)$ is the cumulative probability of being within $|\delta_i/\sigma_{68.3}|$ standard deviations of the mean, $\delta_i$ is the $i$th sorted deviation, and $0<\Delta i<1$ is the bin center; we set $\Delta i=0.683$ to yield intuitive results for the case of $N\rightarrow 1$ and $\mu$ being known \textit{a priori} (see \S 6 of RCR1). 

Solving Equation \ref{68.3_equation1} for $\delta_i$ gives 
\begin{equation}
\delta_i = \sigma_{68.3}\left[\sqrt{2}\mathrm{erf}^{-1}\left(\frac{i-0.317}{N}\right)\right],
\end{equation}
a linear equation for $\delta_i(\sigma_{68.3})$ with slope $\sqrt{2}\mathrm{erf}^{-1}[(i-0.317)/N]$ and intercept zero. Note that if a fraction of the sample is contaminated, the shape of this distribution changes: the curve of $\delta_i$ vs. $\sqrt{2}\mathrm{erf}^{-1}[(i-0.317)/N]$ breaks upwards (\S 2.2, RCR1). As such, for the third method of estimating $\sigma_{68.3}$, we fit a \textit{broken line} of intercept zero to this data, and take the slope of the first component to be $\sigma_{68.3}$ (see Appendix A and Figures 2 and 3 of RCR1 for the details). We also note that as described in \S 2.3 of RCR1, we have calibrated all $2\times3$ combinations of these robust methods for improved performance, in order to avoid ``runaway'' over-rejection of measurements\footnote{\label{footnote:correctivefactor}Specifically, we multiply the computed deviations by correction factors that give the best result, on average, dependent on $N$; see Figure 4 of RCR1 for more details.}.

As will be described in \S\ref{sec:rcr}, in order to have an outlier rejection method that is both robust \textit{and} precise, we use CR with the traditional measures of central tendency and sample deviation (mean and standard deviation) \textit{and} CR with more robust measures (median, mode and 68.3-percentile deviation) in sequence. The exact sequencing and robust techniques used are dependent on the shapes of the uncontaminated distribution and that of the contaminants, shown in Figure \ref{fig:flow_techniques}. This way, we have the ``best of both worlds'' of both robustness and precision. Before we introduce the RCR algorithm as a whole, however, we will need to briefly discuss how the above methods can be generalized to other types of data distributions.

\subsubsection{Distributions That Can Be Used With RCR} \label{sec:nongauss}
RCR (and the robust techniques of \S\ref{sec:robust} that it uses) has been configured and thoroughly tested to work with a variety of types of sample distributions. The uncontaminated measurements can be either (1) normally distributed, (2) non-normally, but symmetrically, distributed, i.e.\ ``peaked'' or ``flat-topped''\footnote{``Peaked'' and ``flat-topped'' meaning the exponential power distribution/generalized normal distribution with positive and negative kurtosis, respectively.} (RCR1 \S 3.3.2), or (3) mildly asymmetric (RCR1 \S 3.3.1). The contaminants can be two-sided (completely symmetric; RCR1 \S 3.1), one-sided (asymmetric; RCR1 \S 3.2) or in-between. In turn, different robust techniques will be used for different distribution types, as described in the following section.

\subsection{The RCR Algorithm: Accuracy \textit{and} Precision} \label{sec:rcr}
The accuracy of the more robust measures of central tendency (median and mode) and of sample deviation (68.3-percentile deviation) comes with a tradeoff: a loss of precision (RCR1). The key question here is then whether it is possible to have \textit{both} accuracy and precision. As we presented in \S 4 of RCR1, this \textit{can} be done, by (1) using traditional Chauvenet Rejection (CR), but with our robust improvements (\S \ref{sec:robust}), and then (2) using traditional CR. Traditional CR uses the least-robust mean and standard deviation, but is also the most precise. By performing CR with our robust methods first, we can remove the outliers that would most significantly affect traditional CR, such that afterwards, the usage of traditional CR will be both precise and accurate.

\begin{figure*}
    \begin{center}
    \includegraphics[scale=0.6, angle=0]{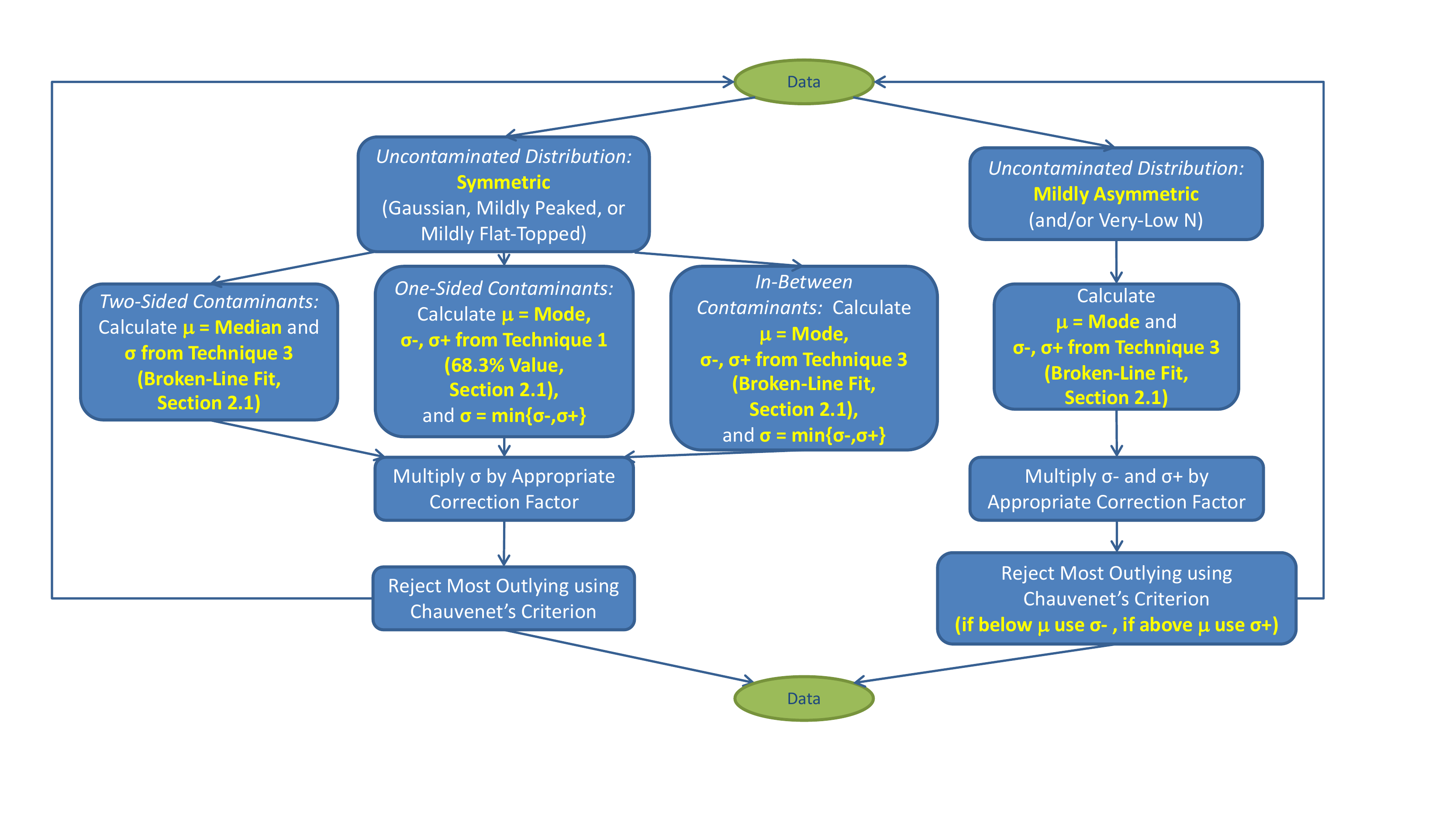}
    \caption{Best-option robust techniques for characterizing different uncontaminated distributions and contaminant types. The most discrepant outlier is rejected each iteration, and one iterates until no outliers remain. $\mu$ and $\sigma$ (or $\sigma_{-}$ and $\sigma_{+}$, one-sided deviations used if the uncontaminated distribution and/or the contaminants are asymmetric) are recalculated after each iteration, and the latter is multiplied by the appropriate correction factor (Footnote \ref{footnote:correctivefactor}) before being used to reject the next outlier.}
    \label{fig:flow_techniques}
    \end{center}
\end{figure*}

Shown in \S 3 of RCR1, we have determined the optimal choices of robust techniques used for CR, dependent on the user-specified type of sample distribution (considering both the contaminated and the uncontaminated distributions). These techniques are summarized in Figure \ref{fig:flow_techniques}, and they were chosen after testing RCR on simulated data sets of different sizes with a variety of distributions, contaminant fractions and strengths, and seeing which of the techniques best recovered the information characterizing the underlying ``true'' distribution (see Figs. 6 - 9 and 13 - 16 of RCR1 for more information).

We have seen that in all cases, using such robust techniques with CR followed by traditional CR gives vastly improved precision, with only small compromises in accuracy (due to not all outliers being rejected via the robust techniques before traditional CR is applied). As such, to further improve our algorithm, we sequence (1) the best-option robust technique with (2) the most precise robust technique (median + $\sigma_{68.3}$ technique) and 3) traditional CR, in order to remove as many outliers as possible before the third step.
While this algorithm is both accurate and precise, it can be quite computationally intensive, especially for large data sets, as \textit{only one} outlier is rejected per iteration. As such, rejecting outliers \textit{in bulk} per iteration can greatly increase computation speed, while---as we will show in the following section---only resulting in slight loses to precision and accuracy.

\subsubsection{Bulk Rejection} \label{sec:bulk}
\begin{figure}
    \begin{center}
    \includegraphics[scale=0.65,angle=0]{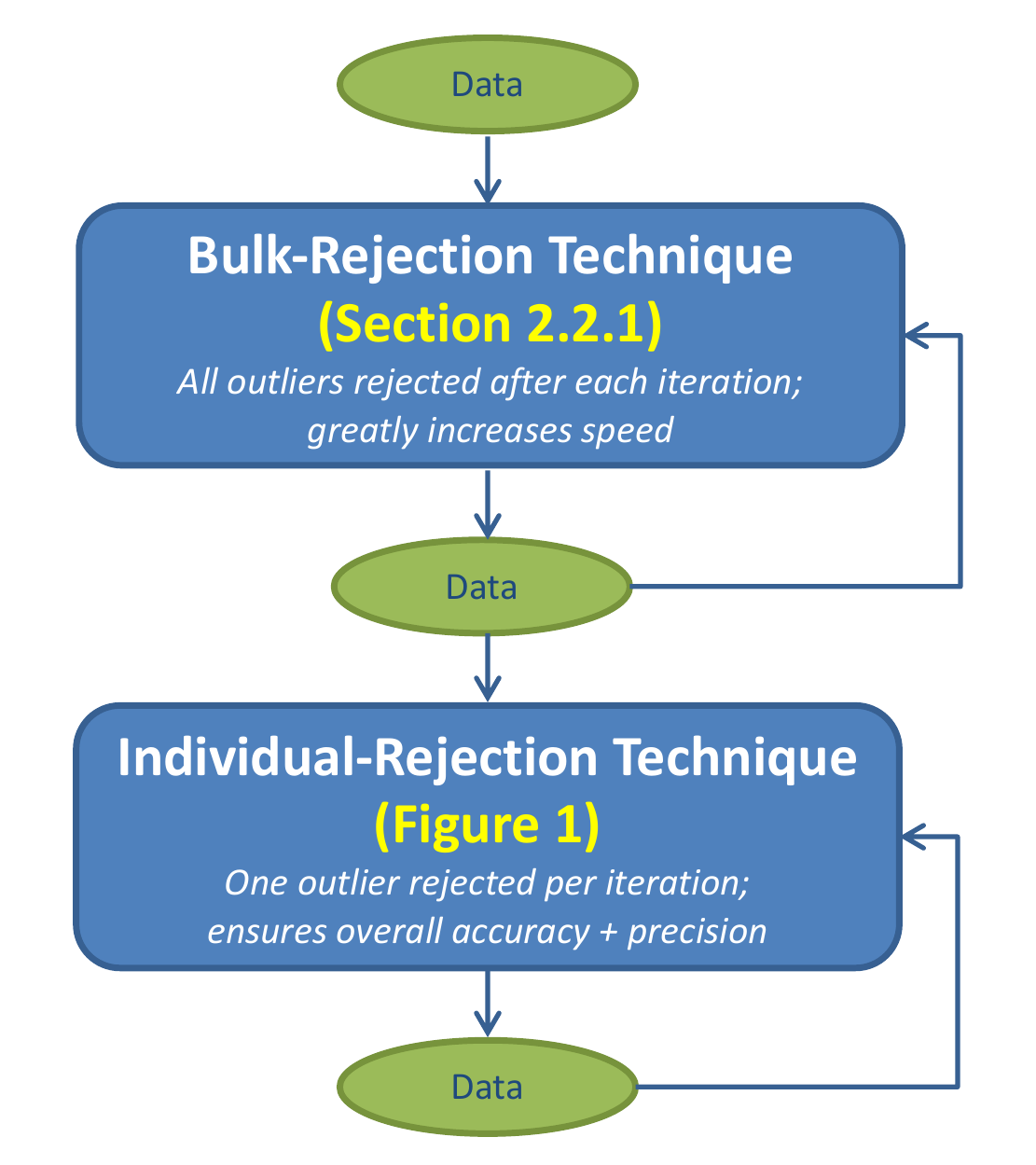}
    \caption{Flowchart of the RCR algorithm. The first step is bulk rejection (\S \ref{sec:bulk}), in which all outliers are rejected each iteration, and one iterates until no more outliers are identified, providing a great boost of efficiency. Central tendency ($\mu$) and deviation ($\sigma$) parameters are recalculated after each iteration to reject more outliers. The second step is our individual-rejection algorithm (Figure \ref{fig:flow_techniques}), which ensures both accuracy and precision (\S \ref{sec:rcr}).}  
    \label{fig:flow_withbulk}
    \end{center}
\end{figure}

The RCR algorithm rejects only one outlier at a time, recomputing central tendency and deviation parameters $\mu$ and $\sigma$ (robust or otherwise), respectively, after every rejection. Computationally this can be prohibitive, especially for large data sets. As such, we need to explore the effectiveness of \textit{bulk rejection}: rejecting all measurements that meet Chauvenet's criterion (Equation \ref{eq:chauvenet}) each iteration, but now only recomputing $\mu$ and $\sigma$ once per iteration instead of once per rejection\footnote{One caveat of bulk rejection is that it only works if the deviation of the uncontaminated distribution is \textit{never} significantly underestimated; if this happens, serious over-rejection of uncontaminated measurements can occur. However, while the causes of such underestimation are explored in \S5 of RCR1, we have found a solution that is effective is all cases, presented in this same section.}. To implement bulk rejection, we use the same $\mu$-measurement techniques as we use for individual rejection (Figure \ref{fig:flow_techniques}). Then, once bulk rejection is completed we continue on to individual rejection. The complete RCR algorithm is shown as a flowchart in Figure \ref{fig:flow_withbulk}.

As shown in RCR1, we have tested the usage of RCR with and without bulk rejection over a large range of data sets, with different distribution types, contaminant fractions and strengths, and sample sizes (Figs. 20 - 27 of RCR1). Overall, we have found that this bulk \textit{pre}-rejection greatly increases the speed (Table 1 of RCR1) of the subsequent individual rejection cycle---due to the massively reduced data set size---while the individual rejection ensures both accuracy and precision of the entire process (\S \ref{sec:robust}).

\subsection{Model Fitting with RCR} \label{sec:functional}
So far, we have only considered using RCR to reject outliers from a 1-dimensional sample $\left\{y_i\right\}_{i=1}^N$, distributed about a single, parameterized ``true'' value $y$. However, RCR can also be used on measurements distributed about some $M$-parameter, $n$-dimensional \text{model} $y\left(\bm{x}|\bm{\theta}\right)$, where $\bm{x}$ is the set of the model's independent variable(s) ($\bm{x}\in\mathbb{R}^n$), and $\bm{\theta}$ are the model's parameters ($\bm{\theta}\in\mathbb{R}^M$). Just as we used the increasingly robust techniques of central tendency and sample deviation (\S\ref{sec:robust}) to fit the 1D $y$ to measurements, we can introduce new, similarly increasingly robust ways to fit this $n$D $y\left(\bm{x}|\bm{\theta}\right)$ to measurements, as follows.

Consider some $n$-dimensional data set $\left\{\left(\bm{x}_i, y_i\right)\right\}_{i=1}^N$. Commonly, we have that the intrinsic uncertainties/error bars of the independent variable(s) are all approximately zero, i.e.\ $\left\{\sigma_{x_{l,i}}\right\}_{l=1}^n = 0 \,\, \forall i$, and that the error bars of the dependent variable are approximately symmetric, i.e.\ $\sigma_{y+,i} \approx \sigma_{y-,i} \equiv \sigma_{y,i} \,\, \forall i$. In the more general case, we can allow for non-zero error bars along the independent variable(s) and asymmetric error bars along both independent and dependent variables, as well as extrinsic uncertainty/sample variance that is not fully accounted for by the error bars, but this makes the situation significantly more involved, e.g.\ \citet{trotter}, \citet{konz} and \citet{TRKIapjs}. For RCR, we only consider the more simple case, but approaching the general case is a possible future endeavor.

Models are often fit to data using a maximum likelihood estimation (MLE) approach, i.e.\ determining the model parameters $\bm{\theta}$ that maximize some likelihood function $\mathcal{L}$ that describes the likelihood that some set of model parameters can generate the data set\footnote{Alternatively, a maximum \textit{a posteriori} approach (MAP) can be taken if prior information is known about the distribution(s) of the model parameters.}. In our aforementioned simplified measurement uncertainty case, the likelihood is simply found with $\mathcal{L}\propto\exp\left(-\chi^2/2\right)$, where
\begin{equation}
	\chi^2 \equiv \sum_i^N \left[\frac{y_i - y\left(\bm{x}_i|\bm{\theta}\right)}{\sigma_{y,i}}\right]^2 \approx N - M.
\end{equation}
As described in \S8.1 of RCR1, this MLE approach is effectively a generalization of the mean within parameter space, and as such is \textit{not} robust. While there are a number of methods for modifying likelihood functions to account for/model outliers, e.g.\ \citet{hogg2010data}, these often make prohibitive \textit{a priori} assumptions about the distribution of the outliers itself. 
To create a more-robust alternative to MLE, we apply the RCR algorithm, except replacing the weighted mean, median and mode with parameter space-based generalized versions of each, in order to maintain their differing levels of robustness and precision. As described, we generalize the weighted mean with MLE, and take the following novel approach to generalizing the median and the mode. 

Consider that for some $M$-parameter model, any combination of $M$ measurements can uniquely determine a set of parameter values $\bm{\theta}$\footnote{In practice, we find best fit parameters for arbitrary models with the nonlinear least-squares Gauss-Newton (GN) algorithm, e.g.\ \citet{floudas2008encyclopedia}, which requires that the user supply first derivatives of $y\left(\bm{x}|\bm{\theta}\right)$ with respect to $\bm{\theta}$. (We use a slightly modified version of the GN algorithm; see footnote 19 of RCR1, page 40).}. If we do this for all possible $M$-combinations of data points\footnote{Or for as large of a randomly drawn (but without repetitions) subset of these as is computationally reasonable; specifically, we switch over to random draws, where each measurement is drawn in proportion to its weight, when ${N\choose M}>20,000$.}, we will obtain a collection of possible sets of parameters $\left\{\bm{\theta}_k\right\}_{k=1}^{{N\choose M}}$, weighted according to the methodology described in the following paragraph\footnote{See Figure 40 of RCR1 for an example of the usefulness of this weighting. }. We then generalize the weighted median and mode by taking the $M$-dimensional weighted median and mode of $\left\{\bm{\theta}_k\right\}$, respectively, which we define in RCR1 \S 8.1.

We estimate the uncertainties $\bm{\sigma}_\theta\equiv\left\{\sigma_{\theta_j}\right\}_{j=1}^M$ for a single set of fitted parameters $\bm{\theta}\equiv\left\{\theta_j\right\}_{j=1}^M$ using the Jacobian matrix $\mathcal{J}$ of the model 
and the error bars $\left\{\sigma_{y,j}\right\}_{j=1}^M$ of the data points used to perform the fit (see RCR1 \S 8.2.2 for details), so that as shown in \S 8.2.1 of RCR1, the fitted parameter values can be weighted (such that there will be relative weighting between different fits $\bm{\theta}_k$) according to $\bm{w}_\theta\equiv\left\{w_{\theta_j}\propto\sigma_{\theta_j}^{-2}\right\}_{j=1}^M$\footnote{All of this assuming constant RMS scatter $\sigma_y\left(\left\{\bm{x}_i\right\}\right)$ throughout the data set; a common counter-example of this (a logarithmic model) is treated in \S 8.2.3 of RCR1.}.  
We also allow for the usage of (built-in or custom) model parameter prior probability distributions $p(\bm{\theta})$, which affect the weighting $\bm{w}_\theta$ of possible sets of model parameters.

To summarize, in this model-fitting case, the RCR algorithm is applied as before for the one-dimensional case, with the key exception that we no longer use the (weighted) measures of central tendency to establish baseline \textit{values} from which corresponding deviations can be determined. Instead, we use our generalizations of these to determine baseline \textit{model functions} of corresponding robustness and precision, and determine deviations from these functions \footnote{One consideration that we make here is that because more model parameters means more degrees of freedom (DOF), we modify the corrective factors used on the computed deviations (Footnote \ref{footnote:correctivefactor}) to account for the fact that more DOF creates artificially smaller deviations for the same number of measurements. (see footnote 15, pg.\ 37 of RCR1 for details).}. Overall, this is a novel approach to regression, which we will further explore in \citet{PSR}. As will be shown in the following section, we have found that this approach can be very effective for recovering the underlying models for even the most contaminated of data sets.


\section{Results and Examples} \label{sec:results}

Originally implemented in C++ and now ported to Python, the RCR algorithm has been intensely tested and simulated, especially in the field of radio astronomy, where observations are often heavily contaminated. In this section, we will explore a few notable examples that showcase different usages of RCR, including single-value rejection, rejecting outliers while model fitting, aperture photometry, and single dish radio astronomy, from RCR1 and other works.

\subsection{Example: One-Dimensional Outliers} \label{sec:results_value}
\begin{figure}
    \begin{center}
    \includegraphics[scale=0.75,angle=0]{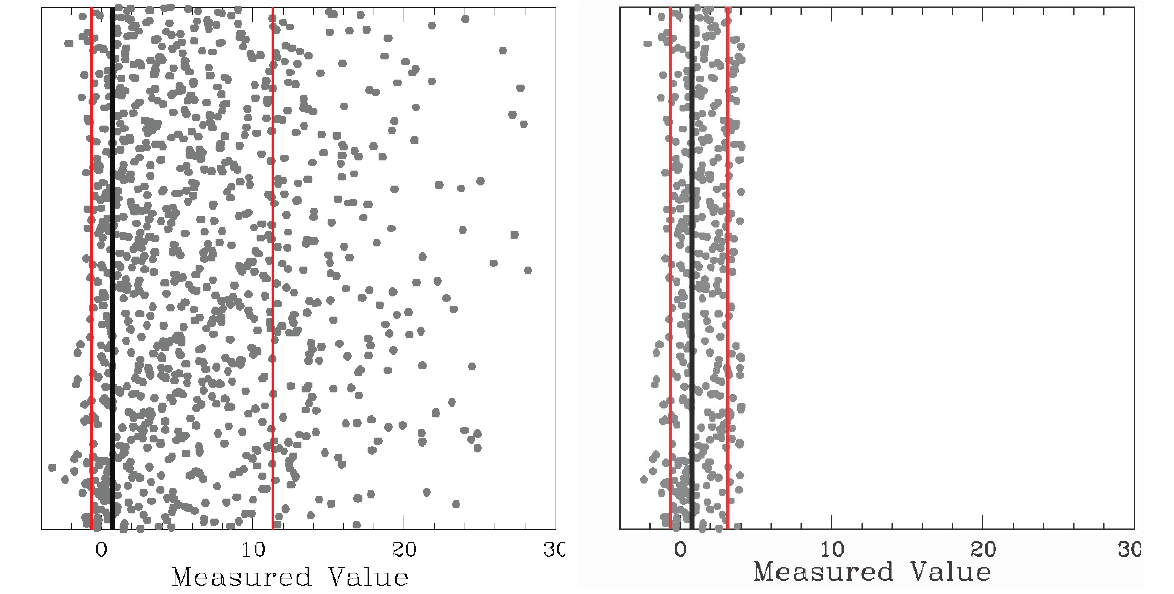}
    \caption{From Figs. 11 and 12 of RCR1: a demonstration of the usage of RCR on a heavily-contaminated sample. \textbf{Left:} A data set of 1000 measurements, with a fraction of 0.15 sampled from a Gaussian distribution with $\mu_1=0$ and standard deviation $\sigma_1=1$ (uncontaminated measurements), and the remaining fraction of 0.85 sampled from the positive side of a Gaussian with $\mu_2=0$ and $\sigma_2=10$, representing one-sided contaminants/outliers. In {\color{black}\textbf{black}} and {\color{black} \textbf{red}} are the best-option robust statistics for this type of distribution (symmetric uncontaminated distribution and one-sided contaminates), the mode and the asymmetric 68.3-percentile deviation (technique 1), respectively (see Fig.\ \ref{fig:flow_techniques} and \S \ref{sec:robust}). \textbf{Right:} The data set after performing RCR with these robust techniques. Note that the correct $\mu\simeq0$ and $\sigma\simeq 1$ is well recovered, even in the face of high contamination.}
    \label{fig:example_value}
    \end{center}
\end{figure}

To begin, see Figure \ref{fig:example_value} for an example of the power of RCR (\S\ref{sec:rcr}) for rejecting outliers while keeping the uncontaminated data largely untouched, on a heavily (85\%) contaminated one-dimensional data set. Here, we present a Gaussian uncontaminated distribution with outliers drawn from the positive side of a much wider Gaussian\footnote{RCR1 explores the same for other, lower contaminant fractions, in Figs. 11 and 12. Unsurprisingly, RCR performs even better with these lower fractions.}. As shown, even in the face of the outlier majority, the underlying ``true'' distribution can be well-recovered with RCR.

\subsection{Example: Model Fitting with Outliers} \label{sec:results_functional}
\begin{figure}
    \begin{center}
    \includegraphics[scale=0.7, angle=0]{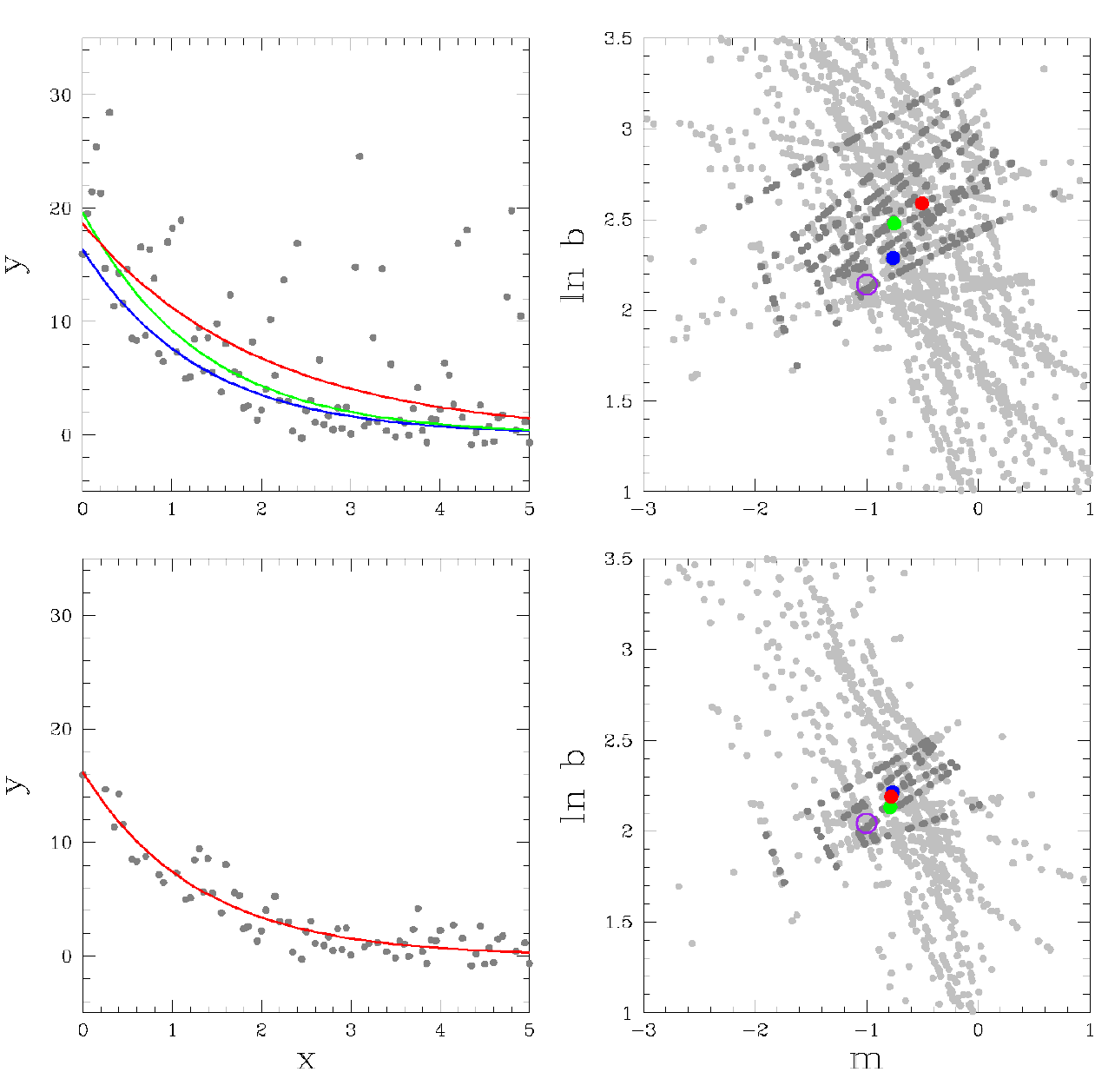}
    \caption{From Figs. 36 and 37 of RCR1: a demonstration of the usage of RCR with model fitting, and a comparison of the various robust techniques used to define a best fit. \textbf{Top Left:} A contaminated data set of 101 measurements, with half sampled from a Gaussian distribution of mean $y(x)=10e^{-(x-0.5)}$ and standard deviation $1$ (uncontaminated measurements), and the remaining half sampled from the positive side of a Gaussian distribution with $\mu=0$ and $\sigma=10$ and added to the uncontaminated measurements, representing one-sided contaminants/outliers. \textbf{Top Right:} Possible model solutions $\left\{\bm{\theta}_k\right\}_{k=1}^{{101\choose 2}}$ calculated from each pair of measurements for the model $y\left[x|\bm{\theta}=(b,m)\right] = be^{m(x-\bar{x})}$ (see \S \ref{sec:functional}), with $\bar{x}$ a fixed value computed to minimize the correlation between $\ln b$ and $m$ (see \S 8.3.5 and Fig.\ 36 of RCR1). Parameter values within each possible model solution are weighted (\S\ref{sec:functional}), with darker points corresponding to models where the product of the weights are in the top 50\%. The {\color{black} \textbf{purple}} circle corresponds to the ``true'' underlying model, the {\color{black} \textbf{red}} circle corresponds to the maximum-likelihood best fit, the {\color{black} \textbf{green}} circle corresponds to the weighted median of $\left\{\bm{\theta}_k\right\}$, and the {\color{black} \textbf{blue}} circle corresponds to the weighted mode. The contaminants bias the calculated models toward shallower slopes and higher normalizations, which is most successfully overcome by the weighted mode technique. \textbf{Bottom:} Same as above, but now after applying RCR with best-option robust techniques  (Fig.\ \ref{fig:flow_techniques}), with rejected model solutions and measurements removed. The final best fit model is shown in {\color{black} \textbf{red}}.}
    \label{fig:example_functional}
    \end{center}
\end{figure}

One-dimensional, or ``single-value'' RCR is quite powerful, but the model-fitting generalization of RCR (\S\ref{sec:functional}), or ``functional'' RCR, has an even wider range of applications, as it can reject outliers about some baseline $n$-dimensional model function (instead of a one-dimensional value), while also fitting that function to the uncontaminated data. RCR1 presented a wide range of examples of the usage of functional RCR to fit different models to data (Figs. 34 -- 42 of RCR1), including linear, quadratic, sinusoidal and exponential models, covering a range of data set sizes and robust techniques. For brevity, we will present just one of these examples in Figure \ref{fig:example_functional}, where we fit an exponential model to a data set that has uncontaminated measurements about an exponential curve, with one-sided contaminants added to 50\% of the measurements. However, as shown in Figs. 36 and 37 of RCR1, when the same procedure is performed with 85\% contaminants, RCR fails; this is because, as we will examine in \S\ref{sec:discussion}, introducing more free parameters to the model decreases the robustness of RCR.

\subsection{Example: Aperture Photometry} \label{sec:results_aperture}
Because of the practically inescapable presence of RFI (radio frequency interference), outliers/contaminants are largely unavoidable when taking observations in radio astronomy. In \S\ref{sec:results_radio}, we will show how RCR is used throughout a complex single-dish radio mapping suite for a number of purposes, including contaminant elimination, background subtraction, and other processing tasks. To begin with, however, we will show a more simple use case of RCR: aperture photometry of the primary radio source Cassiopeia A, or Cas A, from \S 7 of RCR1. 

\begin{figure}
    \begin{center}
    \includegraphics[scale=0.35, angle=0]{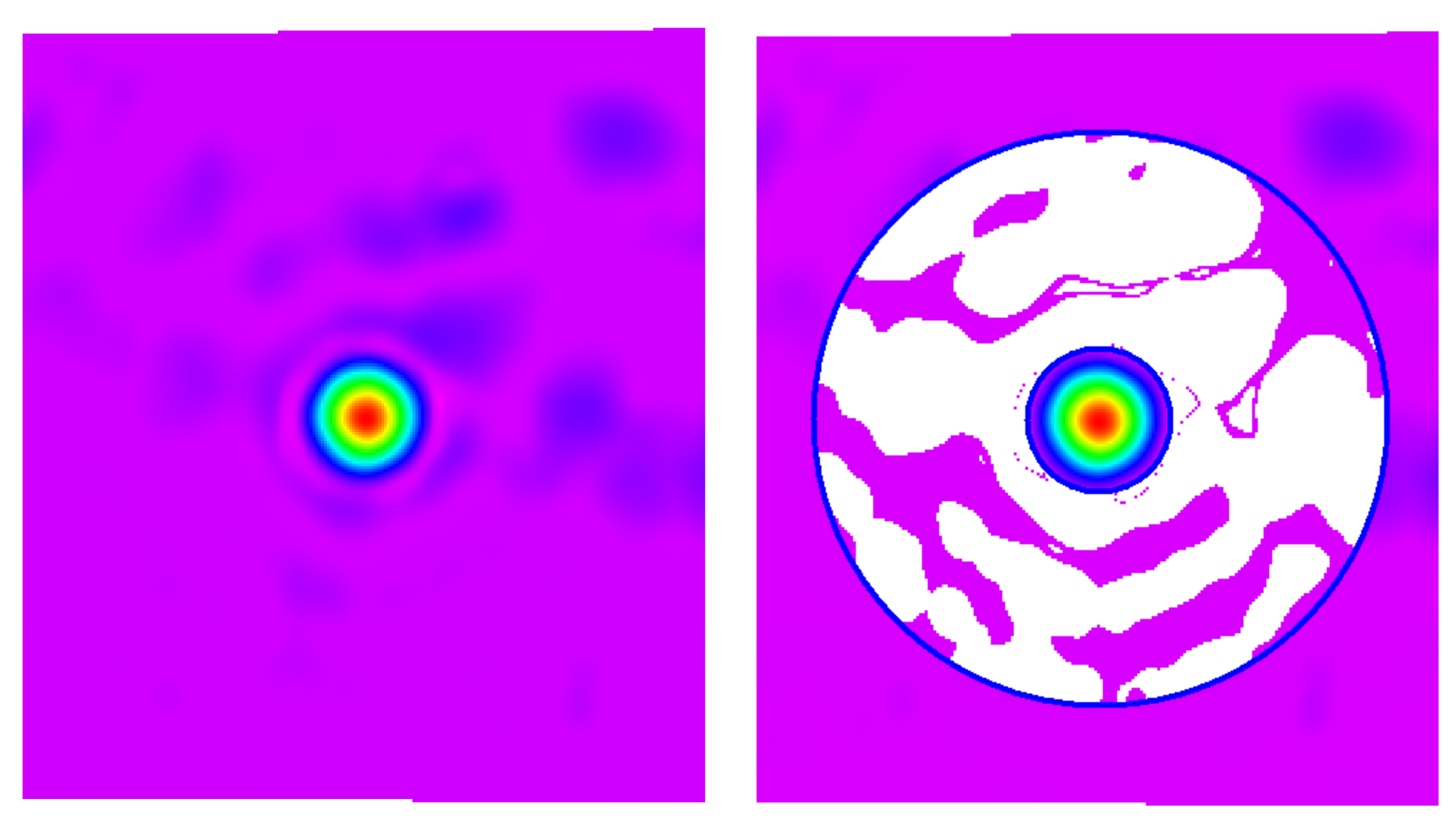}
    \caption{\textbf{Left:} L band image of Cas A, made with Green Bank Observatory's 20-meter diameter telescope, following background subtraction, time-delay correction, RFI cleaning, and surface modeling (in order to fill in gaps between signal measurements without additional blurring). \textbf{Right:} The same image after applying RCR to remove contaminated pixels (i.e.\ from other sources, Airy rings, diffraction spikes, etc.) within an annulus in which we measure the background level. We assume that the contaminants are in-between one-sided and symmetric, e.g.\ there may be some negative contaminants.}
    \label{fig:example_aperture}
    \end{center}
\end{figure}

We observed Cas A in the L band (1.4 GHz) with Green Bank Observatory's 20-meter diameter telescope, which we show the final image of in Figure \ref{fig:example_aperture}, left, after applying various image-processing methods (see caption of Fig.\ \ref{fig:example_aperture}). Assuming a contaminant distribution in-between the one-sided and symmetric case\footnote{Assuming one-sided contaminants gave about the same results as the in-between case (see RCR1 \S 7).} (i.e.\ there are some negative contaminants), we performed RCR with the corresponding rejection technique (Fig.\ \ref{fig:flow_techniques} and \S\ref{sec:rcr}) preceded by bulk rejection (\S\ref{sec:bulk}), to excise the contaminant pixels within a chosen annulus; the resulting image is given in Figure \ref{fig:example_aperture}, right. Numerically, this resulted in a post-rejection background level of $-0.00003 \pm 0.00045$ (gain-calibration units), a massive improvement over the pre-rejection value of $0.023\pm 0.040$, as well as the traditional Chauvenet rejection result of $0.022\pm 0.036$.

\subsection{Example: Single-Dish Radio Mapping Algorithms} \label{sec:results_radio}
The Skynet Robotic Telescope Network is a global network of fully automated volunteer telescopes that are scheduled through a simple, but feature-rich web interface\footnote{\url{https://skynet.unc.edu}}, that has been used for a wide range of scientific and civilian applications. While Skynet originally only included optical telescopes, in the past few years we have been integrating Green Bank Observatory's aforementioned 20-meter radio telescope into the network. As part of this, we have been steadily incorporating RCR into Skynet's image-processing library, beginning with our suite of single-dish radio mapping algorithms, Radio Cartographer (\citet{martin2020}). 

Described in \citet{radioI}, the Radio Cartographer pipeline uses RCR heavily: (1) to eliminate contaminants during gain calibration; (2) to measure the noise level of the data along every scan, and to aid in background subtraction along the scans as a function of time; (3) to combine locally fitted, background-level models into global models for background subtraction along each scan; (4) to excise contaminants if signal and telescope-position clocks must be synchronized post facto from the background-subtracted data; (5) to measure the noise level of the background-subtracted data across each scan, and aid in RFI cleaning over time; and (6) to combine locally fitted models of the background-subtracted, time-delay corrected, RFI-cleaned signal into a global model that describes the observation as a whole\footnote{Following the creation of this global model, we also fill in gaps between the signal measurements by fitting a ``surface'' to the model, in order to produce a final image.}.
\begin{figure}
    \begin{center}
    \includegraphics[scale=0.35, angle=0]{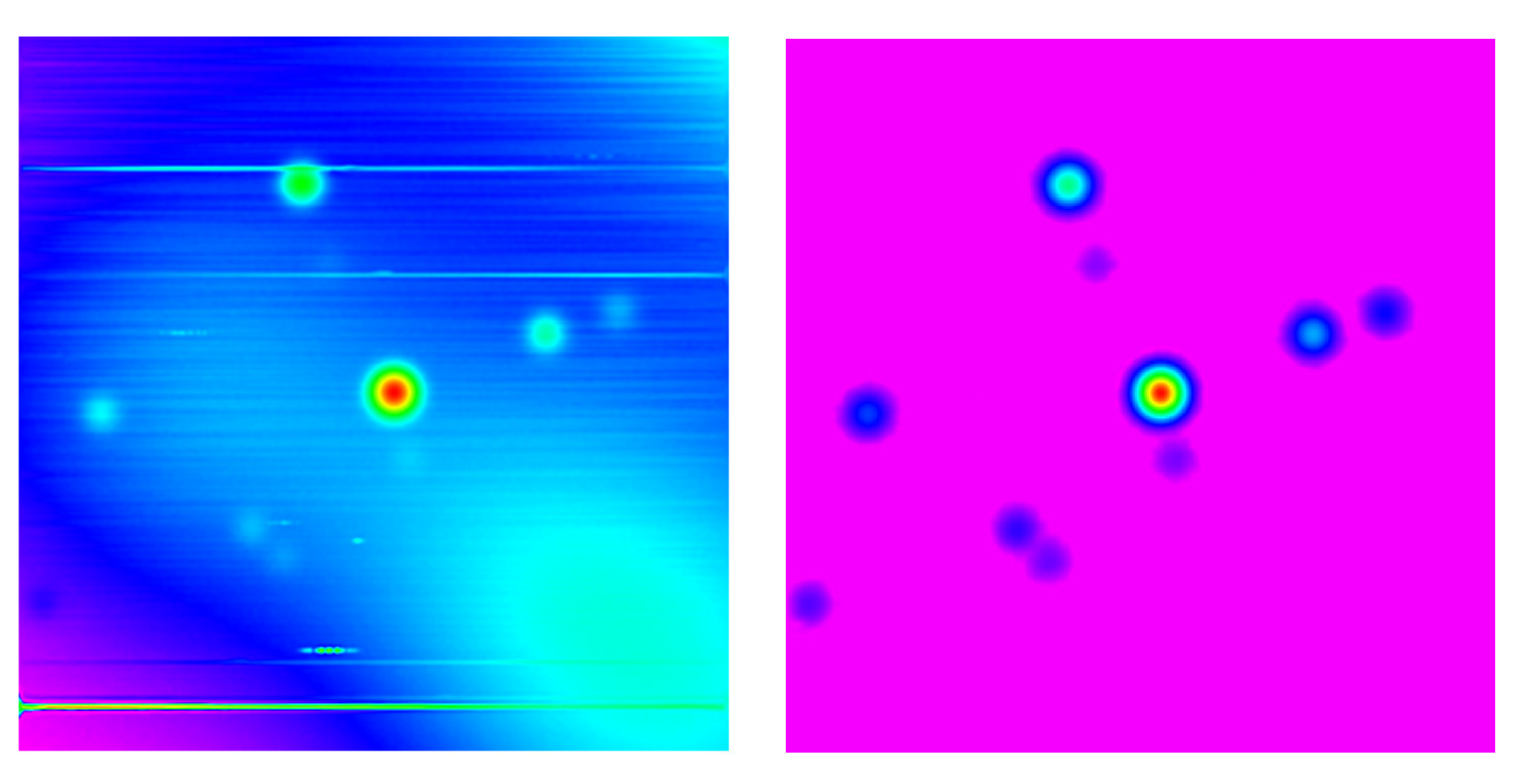}
    \caption{\textbf{Left:} From \citet{radioI}, simulated raw (interpolated) radio data that includes global noise, point sources, short and long duration radio frequency interference (RFI), large scale astronomical and elevation-dependent signal, and en-route drift. \textbf{Right:} the same (interpolated) data after processing with Radio Cartographer, using algorithms built with RCR (see \S\ref{sec:results_radio}).}
    \label{fig:example_radio}
    \end{center}
\end{figure}
In Figure \ref{fig:example_radio}, we show an example of the usage of this pipeline to successfully remove outliers from a simulated radio image that includes a great variety of contaminants. 

Lastly, to provide another example of RCR's successful usage, we reference \citet{trotter2017casA}, who made many measurements of Cas A with Green Bank Observatory's 40-foot and 20-meter telescopes over the course of decades in order to constrain its fading. In this work, RCR was used to reject measurements that were outlying due to variations in the receiver's gain between the primary and calibration observations, which was necessitated over traditional Chauvenet rejection and/or sigma clipping due to a sample contaminant percentage of up to 35\%.


\section{Discussion} \label{sec:discussion}

RCR is a very powerful tool for working with outliers, but a few considerations need to be made before applying it to any sort of data set. The core strength of the algorithm is its ability to maintain both precision and accuracy when rejecting outliers, i.e.\ the estimates of the statistics of the uncontaminated distribution are minimally affected by the outliers. However, this difference is really only noticeable for samples with a moderate to high level of contamination: if the vast majority of your data set is non-outlying, or if there are a few, very obvious outliers that can be pre-removed, then this bias on the uncontaminated statistics from the outliers will be negligible. In such cases, we suggest just sticking with traditional Chauvenet rejection.

Another thing to keep in mind before using RCR is the symmetry of the uncontaminated distribution. As discussed in \ref{sec:nongauss}, although any sort of contaminants---symmetric, one-sided or in-between---can be cleaned via RCR, the \textit{uncontaminated} distribution needs to be fairly symmetric, at most \textit{mildly} asymmetric for RCR to work properly. In particular, the removal of one-sided or in-between contaminants will be especially powerful if this assumption of symmetry holds true, although we have found RCR to still be quite useful in the case of asymmetry.

The usage of RCR for one-dimensional outlier rejection is fairly straightforward. However, if using RCR to fit $n$-dimensional \textit{models} to contaminated data (\ref{sec:functional}) there are further considerations that need to be made, due to the increased generality and possible scope of this application. 

As we saw in \S\ref{sec:results_functional}/Fig.\ \ref{fig:example_functional}, a fundamental limitation of the model-fitting form of RCR is that the algorithm is less robust, i.e.\ more sensitive to outliers as $M$, the number of parameters of the model, increases. To see this, consider that if a fraction $(1-f)$ of $N$ measurements are uncontaminated, a smaller fraction $(1-f)^M$ of the ${N\choose M}$ model solutions are uncontaminated. As such, the higher the dimensionality of the model, the harder it will be for the generalized parameter-space methods of central tendency to ``latch'' on to a desirable solution. In other words, the higher the $M$, the lower the $f$ beyond which RCR fails: the ``curse of dimensionality'' is unavoidable.



Next, let's consider the case of measurements that yield \textit{multiple} model solutions; for example, given a periodic model such as $y(x|b,m,x_0)=b\sin m(x-x_0)$, any combination of $M$-measurements has a countably infinite number of model solutions. In this case, we advise mapping such parameter solutions to a chosen simplest form within the model function definition. For example, this sinusoidal model could be defined to (1) map $m \rightarrow -m$ and $b\rightarrow -b$ if $m< 0$, (2) map $x_0\rightarrow x_0-\frac{2\pi x_0}{m|x_0|}\mathrm{floor}\left(\frac{m|x_0|}{2\pi}\right)$ if $m|x_0|\geq2\pi$, and (3) map $x_0\rightarrow x_0-\frac{\pi x_0}{m|x_0|}$ and $b\rightarrow-b$ if $m|x_0|\geq\pi$. We demonstrated using RCR on such a model with various contaminant fractions in Figures 38-39 of RCR1.

The primary purpose of this work is to introduce RCR in the form of an open source, easy-to-use, professional-quality published Python package. Whereas the original C++ implementation of RCR1 was more ``science code'', the new Python implementation is much more straightforward, well-documented and easier to use and install (see Appendix 1), while also retaining the high speed of the C++ source (thanks to \texttt{pybind11}\footnote{\url{https://github.com/pybind/pybind11}}). While the C++ implementation had a fairly bare-bones pdf documentation, the documentation that we've built for the Python RCR package (see Appendix 2) is fully interactive, includes tutorials for essentially all of the main functionality of RCR, and possesses a full API reference that is automatically updated alongside the main codebase. Finally, we also introduce a web-based calculator through which RCR can easily, but broadly be used (Appendix 3).

\section{Conclusion} \label{sec:conclusion}
Robust Chauvenet Rejection (RCR) is a novel outlier-rejection algorithm originally presented in \citet{maples2018robust}. While the original work provided all of the details regarding the theory, testing and optimization of the algorithm---a conceptual introduction---the purpose of this work is to introduce RCR as an easy-to-use, but advanced collection of software, in the form of a fully-documented, published Python package and a web calculator. With this, we hope for many more to try out RCR for their projects!

\acknowledgments

We would like to thank Michael Maples (UNC Chapel Hill), who helped make RCR a reality by being the original creator of the C++ codebase. We'd also like to thank Dylan Dutton (UNC) and John Martin (UNC/CU Boulder), who contributed to the development of RCR (both codebase and web calculator) in the early stages of the development process. Last but not least, we thank Adam Trotter and Vladimir Kouprianov (UNC Chapel Hill) for their helpful advice and feedback about this work and its accompanying software.

\appendix

\label{app:install}
\section{Installation, Issues and Contributions}

RCR can be easily installed into Python via the Python Package Installer\footnote{\url{https://pip.pypa.io/}}. For Linux or macOS, just use \texttt{python3 -m pip install rcr} in the command line, which will take care of the required packages (really just \texttt{pybind11}) if you do not have them. For Windows, before doing this you'll need to first have Microsoft Visual C++ 14.0, found under the Microsoft Visual C++ Build Tools\footnote{\url{https://visualstudio.microsoft.com/downloads/}}; if that does not work, you may also need the latest Windows SDK\footnote{\url{https://developer.microsoft.com/en-us/windows/downloads/windows-10-sdk/}} (both of these two can be installed via the Visual Studio Installer).

The continual development of RCR can be found on Github\footnote{\url{https://github.com/nickk124/RCR}},
where we push updates and changes to the code, documentation, and web calculator. If you find a bug, or have an idea for a new feature, please create issues or pull requests for us to look at.

\label{app:docs}
\section{Online Documentation}

The documentation\footnote{\url{https://rcr.readthedocs.io/}} for RCR is regularly being updated alongside the codebase. The entire API of the package is documented and annotated, and we include thorough tutorials and examples, from basic installation and one-dimensional outlier rejection to using all of the ``bells and whistles'' of model-fitting/$n$-dimensional RCR. Just like the rest of the codebase, we welcome new ideas for the documentation, via Github.

\label{app:calculator}
\section{RCR Web Calculator}

The final software components accompanying this paper are two web-based calculators. The single-value calculator\footnote{\url{https://skynet.unc.edu/rcr/calculator/value}} can be used to perform one-dimensional
RCR on data that is non-weighted, weighted or with error bars, and we include an interactive plot that allows the user to visualize and interact with the data distribution, as well as change the basis of the data in order to optimize the symmetry of the uncontaminated distribution. 

The functional/model-fitting calculator\footnote{\url{https://skynet.unc.edu/rcr/calculator/functional}} allows the user to use RCR for two-dimensional model-fitting (\S\ref{sec:functional}), again with data that is non-weighted, weighted or with error bars. This calculator includes multiple pre-built model functions, as well as interactive plots of (1) the input data and (2) the fitted model curve alongside the rejected and non-rejected output data. For the former plot, the user can also adjust the bases of the $x$- and $y$-axes of the data such that the RMS scatter of average-weight data is approximately constant along the $x$-axis, in order to best work with RCR.

\bibliography{main}{}
\bibliographystyle{aasjournal}

\end{document}